\documentstyle[prl,aps,twocolumn]{revtex}
\begin{document}
\input{psfig.sty}
\draft

\twocolumn[\hsize\textwidth\columnwidth\hsize\csname
@twocolumnfalse\endcsname

\title
{Energy separation of single-particle and continuum states in a
$S=1/2$ weakly-coupled chains antiferromagnet.}

\author{A. Zheludev$^{(1)}$ \and M. Kenzelmann$^{(2)}$ \and  S.
Raymond$^{(3)}$ \and E. Ressouche$^{(3)}$ \and T. Masuda$^{(4)}$
\and K. Kakurai$^{(5)}$ \and S. Maslov$^{(1)}$ \and I.
Tsukada$^{(4,6)}$ \and  K. Uchinokura$^{(4)}$  and A.
Wildes$^{(7)}$.}

\address{(1) Physics Department, Brookhaven National Laboratory, Upton, NY
11973-5000, USA. (2) Oxford Physics, Clarendon Laboratory, Oxford
OX1 3PU, UK. (3) DRFMC/SPSMS/MDN, CENG, 17 rue des Martyrs, 38054
Grenoble Cedex, France. (4) Department of Applied Physics, The
University of Tokyo, 6th Engineering Bld., 7-3-1 Bunkyo-ku, Tokyo
113-8656, Japan.(5) Neutron Scattering Laboratory, Institute for
Solid State Physics, The University of Tokyo, Tokai, Ibaraki
319-1106, Japan. (6) Present address: Central Research Institute
of Electric Power Industry, 2-11-1, Iwato kita, Komae-shi, Tokyo
201-8511, Japan. (7) Institut Laue-Langevin, Ave. des Martyrs,
Grenoble Cedex 9, France.}

\date{\today}
\maketitle
\begin{abstract}
Inelastic neutron scattering is used to study transverse-polarized
magnetic excitations in the quasi-one-dimensional $S=1/2$
antiferromagnet BaCu$_2$Si$_2$O$_7$, where the saturation value
for the N\'eel order parameter is $m_0=0.12$~$\mu_{\rm B}$ per
spin. At low energies the spectrum is totally dominated by
resolution-limited spin wave-like excitations. An excitation
continuum sets in above a well-defined threshold frequency.
Experimental results are discussed in the context of current
theories for weakly-interacting quantum half-integer spin chains.
\end{abstract}


]

\narrowtext

One of the outstanding problems in quantum magnetism today is that
of dimensional crossover in systems composed of weakly-coupled
spin chains. In particular, for $S=1/2$ Heisenberg
antiferromagnets (AFs), the interplay between classical dynamics
in N\'eel-ordered three-dimensional materials (single-particle
order-parameter excitations, or spin waves) and critical dynamics
of the quantum-disordered $S=1/2$ one-dimensional (1D) model
(2-spinon excitation continuum \cite{Fadeev81,Haldane93}) is not
fully understood. In quasi-1D $S=1/2$ systems, for arbitrary weak
inter-chain interactions, long-range order is restored. However,
according to recent quantum-mechanical
calculations\cite{Schulz96,Essler97}, as long as the 3D order
parameter remains small, the excitation spectrum has a unique {\it
dual} nature. At low frequencies the diffuse continuum is
``cleaned up'' and replaced by long-lived single-particle
``magnon'' excitations. What remains of the critical dynamics in
isolated chains, is now seen at higher energies as a series of
multi-magnon states. The lowest-energy contribution is from
two-magnon processes. At the 1D AF zone-center the continuum is
thus expected to set in at twice the characteristic magnon energy,
and should be separated from the single-particle part of the
spectrum by a ``second gap''.

To date, this separation of single-particle and continuum dynamics
has not been clearly observed experimentally in any quasi-1D
material. Among other technical obstacles, is the limited choice
of suitable model compounds. Most of what is experimentally known
of coupled $S=1/2$ chains comes from neutron scattering studies of
KCuF$_3$ \cite{Satija80,KCUF3,Tennant95,Lake00}. Here the ordered
moment is rather large, making the spin waves very intense at low
energies, and the continuum contribution difficult (though not
impossible) to isolate. Materials like Sr$_2$CuO$_3$
\cite{Kojima97} or SrCuO$_2$\cite{Matsuda97,Zaliznyak00}, on the
other hand, are very good 1D systems, and it is the spin waves
that are hard to identify. Below we report inelastic neutron
scattering studies of spin dynamics in the $S=1/2$ quasi-1D
antiferromagnet BaCu$_2$Si$_2$O$_7$. Using neutron setups with
complimentary resolution characteristics, we clearly observe the
frequency-separation of single-particle and continuum
contributions. In doing so, we pay special attention to the
polarization of magnetic excitations, and compare the results to
existing theoretical models.

The recently characterized BaCu$_2$Si$_2$O$_7$ (orthorhombic space
group $Pnma$, $a=6.862$~\AA, $b=13.178$~\AA, $c=6.897$~\AA )
appears to be an ideal model material for the present study.
Compared to KCuF$_3$, it has a stronger 1D character: with an
in-chain (along the $c$-axis) Heisenberg exchange constant $J=
279$~K, the N\'eel temperature is only $T_{\rm N}=9.2$~K. The
structure of the magnetically ordered phase was previously guessed
from bulk data and an analysis of a {\it single} magnetic Bragg
reflection \cite{Tsukada99}. As part of the present study, we
performed a comprehensive neutron diffraction study of the
magnetic structure  at $T=1.5$~K using the D23 single-crystal
diffractometer at Institut Laue-Langevin (ILL) \cite{elsewhere}.
Overall, 35 inequivalent magnetic Bragg intensities were measured.
The spin arrangement was found to be exactly as proposed in
Ref.~\cite{Tsukada99}.  The obtained value for the ordered moment
$m_0=0.12$~$\mu_{\rm B}$ is much smaller then in KCuF$_3$
($m_0\approx0.5$~$\mu_{\rm B}$), yet large enough to expect a
measurable spin-wave contribution to inelastic scattering.

Low-energy magnetic excitations in BaCu$_2$Si$_2$O$_7$ were
studied in high-resolution inelastic neutron scattering
experiments on a $5\times 5\times 50$~mm$^3$ BaCu$_2$Si$_2$O$_7$
single-crystal sample at the IN14 cold-neutron spectrometer at ILL
\cite{elsewhere}. Measurements were performed in the $(0,k,l)$
scattering plane with a fixed final neutron energy of 3~meV.
Typical constant-$Q$ and constant-$E$ scans collected near the 1D
AF zone-center $l=1$ are shown in Fig.~1. Figure 2 shows a series
of constant-$Q$ scans for different momentum transfers
perpendicular to the chain-axis. The polarization dependence of
the magnetic neutron scattering cross section ensures that within
this range, to a good approximation, only fluctuations of spin
components perpendicular to the chains (and thus the ordered
moment, which is parallel to $c$) are observed. Very similar data
sets (not shown) were collected using a comparable setup on the
TASP cold-neutron 3-axis spectrometer at Paul Scherrer Institut,
Switzerland, for momentum transfers along the $(h,0,1)$ and
$(h,h,1)$ reciprocal-space rods \cite{elsewhere}.
\begin{figure}
\psfig{file=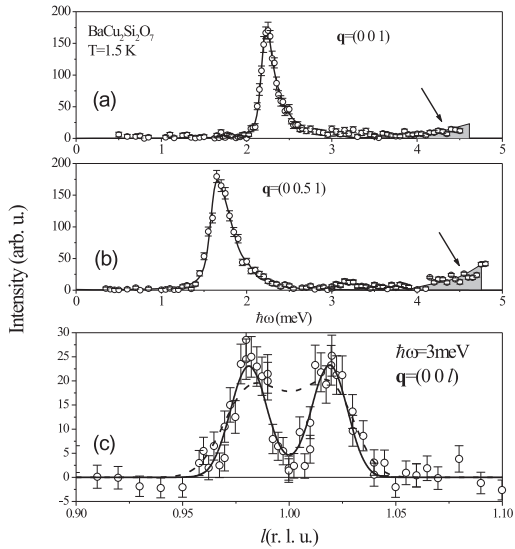,width=3.0in} \caption{Typical constant-$Q$
(a,b) and constant-$E$ (c) scans measured in BaCu$_2$Si$_2$O$_7$
at $T=1.5$~K near the 1D AF zone-center $l=1$ (symbols). Lines are
as described in the text.} \label{fig1}
\end{figure}
The main feature seen in all constant-$Q$ scans is a sharp
asymmetric peak that we attribute to single-magnon excitations.
The intensity onset on the low-energy side is very steep, and
suggests the peaks are resolution-limited. The extended
high-energy ``tail'' is due to a steep dispersion along the chain
axis and a non-zero wave vector resolution in that direction. The
data were analyzed using a single-mode approximation (SMA) cross
section, derived from the quantum chain-Mean Field (chain-MF)
model \cite{Essler97}. The dispersion measured along $(0,k,1)$,
$(h,0,1)$ and $(h,h,1)$, we found that at least three inter-chain
exchange constants are needed, between nearest-neighbor spins
along the $(1,0,0)$, $(0,1,0)$ and $(1,1,0)$ real-space
directions, that we denote as $J_x$, $J_y$ and $J_3$,
respectively. The SMA cross section was convoluted with the
calculated experimental resolution function and used to fit all
measured scans to determine $J_x$, $J_y$ and $J_3$. The in-chain
coupling constant $J$ was fixed at $J=24.1$~meV, as previously
deduced from bulk susceptibility curves \cite{Tsukada99}. An
excellent global fit to all scans is obtained with
$J_x=-0.463(2)$~meV, $J_y=0.161(1)$~meV
 and $2J_3=0.145(1)$~meV. Simulations based on
these values are shown in solid lines in
Figs.~\ref{fig1},\ref{fig2}. The resulting dispersion relation
along the $(0,k,1)$ direction is shown in a solid line in the
$(\hbar\omega)$-$k$ plane in Fig.~\ref{fig2}. Additional analysis
also gives an upper estimate for the intrinsic excitation width:
$0.03$~meV. These results conclusively demonstrate that up to
about 4~meV energy transfer the magnetic excitation spectrum is
fully accounted for by the single-particle picture, in the entire
range of momentum transfers perpendicular to the chain-axis.

\begin{figure}
\psfig{file=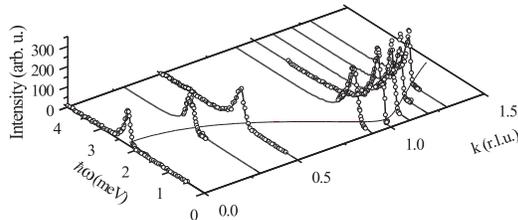,width=2.7in}\caption{A series of constant-$Q$
scans measured in BaCu$_2$Si$_2$O$_7$ at $T=1.5$~K for different
momentum transfers perpendicular to the chain axis. Lines are as
described in the text.} \label{fig2}
\end{figure}
\begin{figure} \psfig{file=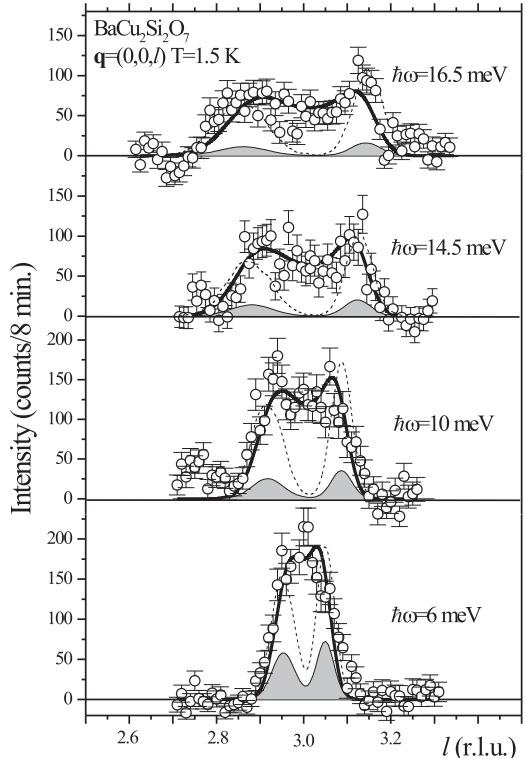,width=2.7in}
 \caption{Constant-$E$ scans
measured in BaCu$_2$Si$_2$O$_7$ at $T=1.5$~K using a
thermal-neutron setup. Lines are fits to the data, as discussed in
the text. Shaded areas are the extrapolated single-mode
contributions.} \label{fig3}
\end{figure}

Extending the cold-neutron study to higher energy transfers was
hindered by certain geometrical constrains imposed by the design
of IN14 spectrometer. Instead, the intermediate-energy spectrum
was studied using the IN22 thermal-neutron spectrometer at ILL
using 14.7~meV final-energy neutrons. Most of the data were
collected in constant-$E$ scans along the chain-axis, in the
vicinity of the $(0,0,3)$ 1D AF zone-center. The background
originating from the empty Al sample-holder was measured
separately for each scan. Typical background-subtracted data sets
are presented in Fig.~\ref{fig3}.
 None of these show the distinct 2-peak structure
as at 3~meV energy transfer (Fig.~1c). To verify that this is not
a resolution effect, we performed a least-squares fit of the SMA
cross section, convoluted with the calculated resolution, to each
measured profile. Very poor fits are obtained, as shown in dashed
lines in Fig.~\ref{fig3}. It is clear that if the observed
intensity was due to a single mode, two separate peaks would have
been easily resolved. For lack of a convenient analytical result
for interacting chains, we fitted the measured scans to the
M\"uller-ansatz continuum form \cite{Muller}, known to work well
for isolated chains. The agreement in this case is much better
(solid lines in Fig.~\ref{fig3}). For contrast, the dashed line in
Fig.~\ref{fig1}c shows the best M\"uller-ansatz fit to the 3~meV
high-resolution const-$E$ scan. From this comparative analysis we
conclude that, unlike at low energies, a large contribution to the
observed scattering at high energies must come from continuum
excitations.

\begin{figure}
\psfig{file=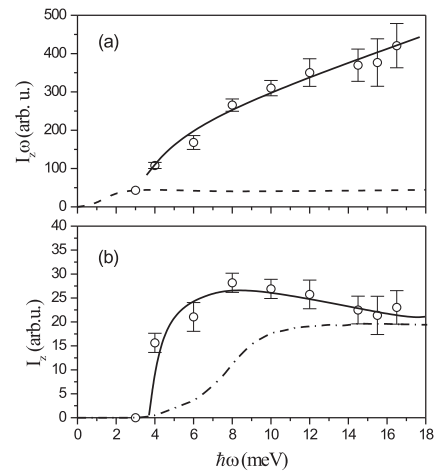,width=2.5in} \caption{ (a) Measured
$q_z$-integrated intensity scaled by energy transfer (symbols).
The dashed line is the extrapolated single-mode contribution. (b)
Estimated $q_z$-integrated intensity of continuum excitations in
BaCu$_2$Si$_2$O$_7$ (symbols).  The dash-dot line is the
calculated spin wave theoretical 3-magnon spectrum. In both panels
the solid line is a guide for the eye.}\label{fig4}
\end{figure}

Additional evidence was obtained in analyzing the measured
intensities. In the chain-MF model the intensity of magnon
scattering has the same $1/\omega$ dependence as for classical
antiferromagnetic spin waves, in full agreement with the bulk of
our cold-neutron data. Having confirmed the single-mode picture at
low energies, we exploited this scaling relation to estimate spin
wave intensities at higher energy transfers. A normalization
factor was obtained by fitting the SMA cross section to the
intensity measured using the thermal setup at $\bbox{q}=(0,0,3)$
between 2 and 3~meV transfer, where, according to our cold-neutron
results, the SMA picture is still valid. The thus calculated
single-mode contributions are represented by shaded areas in
Fig.~\ref{fig3}. Clearly, magnon excitations account for only a
minor fraction of the dynamic susceptibility at high frequencies.
To quantify this behavior, in Fig.~\ref{fig4}a we plot the
measured $q_z$-integrated intensity $I_z$, scaled by energy
transfer $\hbar \omega$, as a function of the latter. The
advantage of this integration procedure that the actual
wave-vector dependence of continuum scattering becomes
unimportant. The expected spin wave contribution is practically
constant above 3~meV in this plot (Fig.~\ref{fig4}a, dashed line).
This follows from that in a fixed-final energy configuration one
directly measures $S(\bbox{q},\omega)$ without any additional
energy-dependent scaling factors ($\lambda/2$ effects in the
monitor are expected to be less than 5\%, thanks to the use of a
thermal guide). In contrast, experimentally, $I_z \omega$
increases substantially with frequency.

The continuum fraction can be isolated by subtracting the
extrapolated SMA part from the experimental data. The result is
shown in Fig.~\ref{fig4}b. The energy resolution being about
1.5~meV, we conclude that the continuum sets in between 3.5 and
5.5 meV energy transfer. We have verified that this result is
quite robust, and remains valid even if we assume that there is a
50\% systematic error in the estimate of the single-mode
contribution. From Fig.~\ref{fig4}b we see that above the
threshold the $q_z$-integrated continuum intensity is only
slightly energy-independent. Looking back at the cold-neutron
measurements, we find that they actually do contain  direct
evidence of continuum scattering as well. Indeed, the increase of
intensity seen around 4.5~meV in Fig.~\ref{fig1} is {\it not}
accounted for by the SMA picture, and thus represents the onset of
the continuum, in full agreement with the thermal-neutron results.

To understand the observed long-lived and continuum excitations,
we shall recall the simple yet profound physical picture provided
by the quantum chain-MF theory \cite{Schulz96,Essler97}. For an
isolated chain in a staggered field, the magnons have an energy
gap $\Delta$ that scales as $H_{\pi}^{2/3}$
\cite{Oshikawa97,Schulz96,Dender97}. In the coupled-chains case,
magnon dispersion perpendicular to the chain axis leads to a
softening of the gap at the 3D zone-center for
transverse-polarized excitations. These gapless modes correspond
to conventional spin waves in the classical spin wave theory
(SWT), and are the long-lived excitations that we see in
BaCu$_2$Si$_2$O$_7$ at low energies. Excitations at exactly
$\Delta$ can be observed in points of reciprocal space where
inter-chain-interactions cancel out at the MF level. In
BaCu$_2$Si$_2$O$_7$ this occurs at $\bbox{q}=(0.5,0.5,1)$. Using
the measured exchange constants we can estimate $\Delta=2.37$~meV.
It is straightforward to verify that the experimental values for
$\Delta$, $J$, $m_0$ and $T_{\rm N}$ satisfy the relations
predicted by the chain-MF model \cite{Schulz96} remarkably well.

A unique quantum-mechanical feature of a half-integer Heisenberg
AF chain in a staggered field, totally absent in conventional SWT,
is a {\it longitudinal} mode, i.e., a magnon excitation polarized
parallel to the direction of static staggered magnetization. The
existence of such an excitation was recently confirmed
experimentally in KCuF$_3$ \cite{Lake00}. Focusing on transverse
spin correlations in the present study, we do not directly observe
this branch in BaCu$_2$Si$_2$O$_7$. It does, nonetheless, play an
important role in the observed transverse {\it continuum}. In the
chain-MF and field-theoretical models for weakly-interacting
chains, low-energy continuum excitations are seen as multi-magnon
states \cite{Essler97}. The existence of a longitudinal mode
allows {\it two}-particle excitations in the transverse channel,
composed of one longitudinal and one transverse magnon. The
threshold for such excitations at the 1D AF zone-center is at
twice the characteristic magnon energy: $\Delta_c=2\Delta$. This
prediction is consistent with our experimental results for
BaCu$_2$Si$_2$O$_7$, where $\Delta_c$ is between 3.5 and 5.5~meV,
and $2\Delta=4.8$~meV.


An alternative model that predicts multi-magnon continua is SWT
with kinematic corrections, which explicitly takes into account
non-linearities in the spin wave Hamiltonian. Continuum
excitations observed in KCuF$_3$ are remarkably well described by
this model \cite{KCUF3,Tennant95}. In BaCu$_2$Si$_2$O$_7$,
however, the ordered moment is much smaller, and the ground state
resembles the N\'{e}el phase, i.e., the starting point of SWT
calculations, considerably less. For this material SWT completely
fails at the quantitative level. For example, the spin wave
correction to sublattice magnetization in BaCu$_2$Si$_2$O$_7$ that
we calculated using the known dispersion bandwidths is well over
100\%. For the {\it transverse}-polarized excitation continuum,
the inadequacy of SWT becomes even more apparent. Indeed, the
longitudinal mode being absent in this model, two-magnon
transverse excitations are prohibited by selection rules. The
lowest-energy contribution is from three-magnon states. The lower
bound of the continuum should thus be at roughly $3\Delta$. The
3-magnon SWT contribution can be exactly evaluated, provided the
spin wave dispersion is known. Following the recipe given in
Ref.~\cite{3magnon}, the spin wave bandwidths measured in
BaCu$_2$Si$_2$O$_7$ were used to calculate the 3-magnon profile at
$\bbox{q}=(0,0,3)$. The result (arbitrary scaling) is shown in
Fig.~\ref{fig4}b in a dash-dot line. The calculated continuum
onset is clearly at a larger energy, by 2 to 3~meV, than observed
experimentally.

In summary,  we find that the transverse excitation spectrum in
BaCu$_2$Si$_2$O$_7$ is divided into two well-defined regions.
Below 4~meV the weight is consolidated into long-lived
single-particle spin wave-like excitations, and no sign of
multi-particle continuum is found. These excitations carry only a
small fraction of the total spectral weight at higher energies,
where an excitation continuum becomes the dominant contribution.
While further studies of the lower continuum bound are needed, the
present results suggest a  threshold between 3.5 and 5.5~meV. For
this strongly 1D material the conventional spin wave- theoretical
description becomes inadequate, even if kinematic corrections are
taken into account. In contrast, the chain-MF model, based on
quantum-mechanical properties of individual chains, is in very
good agreement with experiment.

 We would like to thank Dr. P. B\"{o}ni (PSI
Villigen) for allowing us to refer to the yet unpublished TASP
data, Dr. L.-P. Regnault (CEA Grenoble) for his assistance with
experiments at ILL, Prof. A. Tsvelik, Prof. R. A. Cowley(Oxford
University) and Dr. I. Zaliznyak (BNL) for illuminating
discussions, and Mr. R. Rothe (BNL) for technical support. This
work is supported in part by the U.S. -Japan Cooperative Program
on Neutron Scattering, Grant-in-Aid for COE Research ``SCP coupled
system'' from the Ministry of Education, Science, Sports, and
Culture. Work at Brookhaven National Laboratory was carried out
under Contract No. DE-AC02-98CH10886, Division of Material
Science, U.S.\ Department of Energy. One of the authors (M.~K.) is
supported by a TMR-fellowship from the Swiss National Science
Foundation under contract no. 83EU-053223.


\begin{thebibliography}{10}
\bibitem{elsewhere}
Details of this and other experiments referred to here, will be
reported
  elsewhere.
\end{thebibliography}

\end{document}